\documentclass{aa}

\usepackage{txfonts}
\usepackage{graphicx}
\usepackage{times}
\usepackage{psfig}
\usepackage{longtable}
\usepackage{psfig}
\usepackage[figuresright]{rotating}
\usepackage{subfigure}
\usepackage{aas_macros}
\usepackage{amssymb}
\usepackage{natbib}
\bibpunct{(}{)}{;}{a}{,}{,}
\def\Tiny{\fontsize{6pt}{6pt}\selectfont}

\begin{document}
\title{Star Formation in the Outer Galaxy: Coronal Properties of NGC\,1893\thanks{Tables \ref{tbl:src_properties_main} and \ref{tbl:upl_table}  are only available in electronic form
at the CDS via anonymous ftp to cdsarc.u-strasbg.fr (130.79.128.5)
or via http://cdsweb.u-strasbg.fr/cgi-bin/qcat?J/A+A/}}
\author{M. Caramazza\inst{1,2}
          \and G.}
% \subtitle{I. X-ray properties in a low metallicity environment}
\author{M. Caramazza\inst{1}
          \and  G.
	  Micela\inst{1} \and L. Prisinzano \inst{1} \and S. Sciortino \inst{1} \and F. Damiani \inst{1} \and F. Favata \inst{2} \and J. R. Stauffer \inst{3} \and A. Vallenari \inst{4} \and S. J. Wolk \inst{5} }
\institute{INAF Osservatorio
	  Astronomico di Palermo, Piazza del Parlamento 1, 90134 Palermo, Italy  \and European Space Agency, 8-10 rue Mario Nikis, 75015 Paris, France \and
	  Spitzer Science Center, Caltech 314-6, Pasadena, CA 91125, USA \and  INAF, Osservatorio Astronomico di Padova, Vicolo dell'Osservatorio 5, 35122 Padova, Italy \and Harvard-Smithsonian Center for Astrophysics, 60 Garden Street, Cambridge, MA 02138, USA  \\	  
	  email:mcarama@astropa.unipa.it}
\date{Received ; accepted }

 \abstract
{The outer Galaxy, where the environmental conditions are different from the solar neighbourhood is a laboratory in which it is possible to investigate the dependence of star formation process on the environmental parameters.}
{We investigate the X-ray properties of NGC\,1893, a young cluster { $ (\sim 1-2$ Myr)} in the outer part of the Galaxy ( galactic radius $\geq$ 11 kpc) where we expect differences in the disk evolution and in the mass distribution of the stars, to explore the X-ray emission of its members and compare it with that of young stars in star forming regions  near to the Sun.}
{We analyze 5 deep \textit{Chandra} ACIS-I observations with a total exposure time of 450 ks.  Source events of the 1021 X-ray sources have been
extracted with the IDL-based routine ACIS-Extract. Using spectral fitting and  quantile analysis of X-ray spectra, we derive X-ray luminosities and compare the respective properties of Class II and Class III members. We also evaluate the variability of sources using the Kolmogorov-Smirnov test and we identify flares in the lightcurves. }
{The X-ray luminosity of NGC\,1893 X-ray members is in the range $10^{29.5}-10^{31.5}$ erg s$^{-1}$. Diskless stars are brighter in X-rays than disk-bearing stars, given the same bolometric luminosity. We found that 34\% of the 1021 lightcurves appear variable and that they show $0.16$ flare per source, on average. Comparing our results with those relative to the Orion Nebula Cluster, we find that, accounting for observational biases, the X-ray properties of NGC\,1893 and the Orion ones are very similar. }
 {The X-ray properties in NGC\,1893 are not affected by the environment and the  stellar population in the outer Galaxy may have the same coronal properties of nearby star forming regions. The X-ray luminosity properties and the X-ray luminosity function appear to be universal and can therefore be used for distance estimations  and for determining stellar properties as already suggested by Feigelson and collaborators.}
% 5 {} token are mandatory
\titlerunning{Coronal properties of NGC\,1893}
\maketitle
\section{Introduction}
\label{intro_x}
The formation of stars, the evolution of circumstellar disks and eventually the formation of planets are key research topics in modern astrophysics. So far, several star forming regions of different ages and in different environment conditions have been studied in order to explore the parameter space as well as possible. Although several questions have been answered, some aspects of the star formation mechanism and initial stellar evolution are still obscure.

A key question is whether and how star formation depends on the environmental conditions. Since the Jeans mass depends on gas temperature, chemical abundances, and density \citep[cf.][]{Elmegreen_02, Elmegreen_04}, some dependencies on the physical conditions of star-forming clouds are expected. Moreover, the observed variation in the initial mass function (IMF) at low stellar masses across different environments \citep{Reyle_01, Gould_98} suggests that gravity is not the only important parameter in the cloud fragmentation and subsequent evolution. As a consequence, other processes, e.g. turbulence and/or magnetic field strength in the cloud have been considered \citep{Shu_87,Padoan_99,Maclow_04}.

In this context, a natural way to explore the differences due to the environment is to look at the outer Galaxy. Molecular clouds and very young stellar associations are present also at large distances from the Galactic Center \citep{Snell_02}, implying that star formation is occurring in these regions. Conditions in the outer Milky Way are thought to be less favorable to star formation, when compared with those in the solar neighborhood: the average surface and volume densities of atomic and molecular hydrogen are much smaller than in the solar neighborhood or in the inner Galaxy \citep{Wouterloot_90}, the interstellar radiation field is weaker \citep{Mathis_83}, prominent spiral arms are lacking and there are fewer supernovae to act as external triggers of star formation. Metal content is, on average, smaller \citep{Wilson_92}, decreasing radiative cooling and therefore increasing cloud temperatures and consequently pressure support. Moreover, the pressure of the inter-cloud medium in the outer Galaxy is smaller \citep{Elmegreen_89}.

Given these relevant differences, the study of star forming regions in the outer Galaxy is likely to be helpful to ascertain the role of environmental conditions in the star formation process. First of all, the environmental conditions could affect the evolution of protoplanetary disks: \citet{Yasui_09} measured, in the extreme outer Galaxy, a disk frequency significantly lower than in the inner part of the Milky Way. The eventual dependence of disk lifetimes on metallicity could affect the formation of planets and may give also insight into the ``planet-metallicity correlation'', since stars known to harbor giant planets appear to preferentially be metal rich \citep[][and references therein]{Santos_03,Fischer_05}. Moreover, a recent theoretical works \citep{Ercolano_09} shows that the disk lifetime in regions of different metallicity can give insight into the general mechanism of disk dispersal, allowing to discriminate between the two currently dominant models of disk dispersal: photoevaporation and planet formation.% Indeed, the determination of disk fractions in regions of lower metallicity compared to the solar neighborhood could help to determine the main mechanism responsible for the evolution of protoplanetary discs.

In principle, the metal abundance differences between the inner and the outer Galaxy could also affect the coronal properties. Indeed, the thickness of the outer convection zone in low mass stars is a function of metallicity \citep{Pizzolato_01}. This, in turn, likely changes the dynamo process itself, i.e. the  X-ray emission engine. Moreover, differences in the composition of X-ray emitting plasma may produce other differences: X-ray radiative losses, for example, are dominated by radiation of highly ionized atoms and therefore we may expect changes in the emission measure, as well as, in individual lines. Finally, if there was a difference in the disk fraction, we would expect an indirect effect on the global X-ray emission:  Class II stars have lower X-ray luminosities than Class III  stars \citep{Stelzer_01, Flaccomio_03b, Stassun_04_a, Preibisch_05_b,Flaccomio_06, Telleschi_07} thus a different fraction of disked stars may affect the global X-ray luminosity function. The assumption of a universal X-ray luminosity function has been proposed by \citet{Feigelson_05b} and used to estimate the distance \citep{Kuhn_10} and total populations of several young clusters \citep[e.g.][]{Getman_06, Broos_07}. In this context, the study of a cluster lying in a very different environment could be helpful in demonstrating the universality of X-ray properties of young star forming regions and justify the use of X-ray luminosity functions to determine the properties of young clusters.  

In order to investigate the X-ray properties of a cluster in the outer Galaxy and compare them to similar cluster in the inner regions of the Galaxy, we have identified, as a suitable target, the young ($\sim$ 1-2 Myr) cluster NGC\,1893, whose galactocentric distance is ≥ 11 kpc \citep{Prisinzano_11}. NGC\,1893 is in a spiral arm different than that of the Sun and it is located near the edge of the Galaxy, where the environmental condition are quite different from those of the solar neighborhood. Situated in the Aur OB2
association toward the Galactic anti-center, NGC\,1893 is associated with the HII region IC 410 and the two pennant nebulae, Sim19
and Sim130 \citep{Gaze_52}, known as ``Tadpole''. It contains a group of early-type stars with some molecular clouds but only moderate extinction. Several studies \citep{Vallenari_99, Marco_02, Sharma_07, Caramazza_08, Prisinzano_11} demonstrated that star formation is still ongoing and therefore we can study the X-ray properties of the PMS stars, with particular focus on the differences between the Class II and the Class III stars X-ray emission, given that the disk evolution may depend on the environment \citep{Yasui_09}. The morphology, the age distribution of the cluster and the star formation history have been studied in detail by \citet{Sanz-Forcada_11}. They find that the  cluster has not relaxed yet to a spherical distribution, and it still has different episodes of stellar formation. The analysis of the age and disc frequency of the objects in NGC\,1893, in relation with the massive stars and the nebulae, revealed an ongoing stellar formation close to the dark molecular cloud and the two smaller nebulae Sim 129 and Sim 130 \citep{Sanz-Forcada_11}. While both massive and low mass stars seem to form in the vicinity of the denser molecular cloud, Sim 130 and specially Sim 129 harbour the formation of low mass stars \citep{Sanz-Forcada_11}.
A parameter that is still very uncertain for this cluster is the metallicity. As we discussed in  \citet{Prisinzano_11}, the literature studies about the metallicity of NGC\,1893 have ambiguous results. In a paper focused on the Galactic metallicity, \citet{Rolleston_00} found for 8 NGC\,1893 members a slight indication of under solar abundances, \citet{Daflon_04} show a marginal indication of subsolar metallicity for 2 members of the cluster, but these studies do not show a strong evidence of subsolarmetallicity, therefore, as assumed also in \citet{Prisinzano_11}, we adopt for NGC\,1893 a solar metallicity.

This study is part of a large project aimed at investigating the star formation processes in the outer Galaxy and is based on a large observational campaign on the young cluster NGC\,1893. The first results, concerning the joint Chandra-Spitzer large program ``The Initial Mass Function in the Outer Galaxy: the star forming region NGC\,1893'' (P.I. G. Micela), have been  presented in \citet{Caramazza_08}. In that earlier work we  have  analyzed the {\it Spitzer} IRAC maps of NGC\,1893 joined to the {\it Chandra} ACIS-I long exposure observation, giving the most complete census of the cluster to date. The IRAC observations of NGC\,1893 allowed us to identify the members of the cluster with infrared excesses (Class 0/I and Class II stars), while the ACIS observation was used to select the Class III objects that have already lost their disks and have no prominent infrared excess. Although that first catalog of members was incomplete, the presence of 359 members indicates that NGC\,1893 is quite rich, with intense star forming activity despite the ``unfavorable'' environmental conditions in the outer Milky Way. In \citet{Prisinzano_11}, we continued the study of  the properties of NGC\,1893 by using deep optical and JHK data and compiling a catalog extending from X-rays to NIR data. In this second work, we have assessed the membership status of each star, finding 415 diskless candidate members plus 1061 young stellar objects with a circumstellar disk or Class II candidate members, 125 of which are also H$_\alpha$ emitters. Moreover, optical and NIR photometric properties have been used to evaluate the cluster parameters. Using the diskless candidate members, the cluster distance has been found to be 3.6$\pm$0.2 kpc and the mean interstellar reddening E(B-V)=0.6$\pm$0.1 with evidence of differential reddening across the whole surveyed region. These previous studies show that NGC\,1893 contains a conspicuous population of pre-main sequence stars with a disk fraction of about 70\%, similar to that found in clusters of similar age in the solar neighborhood. This demonstrates that, despite the expected unfavorable conditions for star formation, very rich young clusters can form also in the outer regions of our Galaxy.  
% paper, a conspicuous sample of
% low mass cluster members with circumstellar disk and of more
% evolved diskless candidate members belonging to this region has
% been identified by using Spitzer-IRAC and Chandra X-ray data.
% We present here the results on the membership and cluster
% parameters based on a multiwavelength approach including
% new deep optical and JHK photometry that allow us to derive a
% more complete census of the low mass population in this cluster;
% knowledge of individual members is used to derive cluster parameters
% (reddening and distance) in a more accurate way than
% previously reported in the literature. Detailed studies on the Xray
% coronal properties, disk fraction and evolution and IMF will
% be presented in forthcoming papers.This cluster is a perfect target for our purpose because of its large galactocentric distance ($\eq 11$ Kpc)
% and low metallicity. NGC\,1893 has sub-solar metal content, as reported by \citet{Rolleston_00} from the analysis of spectra of a few B stars. \citet{Daflon_04} confirmed that the abundances of C, N, and O in NGC\,1893 were two to three times lower than in the Sun.

% \citet{Caramazza_08} 
In the present work, as part of the same project,  we present the in-depth analysis of  450 ks long {\it Chandra} ACIS-I observation, already introduced in \citet{Caramazza_08}, focusing on the X-ray properties of the 1021 detected sources in light of the previous results of \citet{Caramazza_08}  and \citet{Prisinzano_11}. To put our work in context, we will compare the present study with the 850 ks long {\it Chandra} ACIS-I observation of the Orion Nebula Cluster (ONC), termed Chandra Orion Ultradeep Project (COUP) \citep{Getman_05}. The COUP with 1616 detected sources provided one of the most  comprehensive dataset so far acquired on the X-ray emission of PMSs. Several studies were conducted on the 1616 COUP detected sources, in order to characterize the X-ray emission and to understand the coronal processes in the PMS stars. For example, \citet{Preibisch_05_b} have analyzed the dependence of the X-ray luminosity on the stellar properties, finding for the Orion stars that  $L_X$ is a function of the mass, \citet{Wolk_05} have studied the variability of the solar mass stars, in order to have indication of the emission of the young Sun, \citep{Favata_05} have investigated the geometry of flare loops, resulting in the first observational indications of magnetic structure connecting the star and the disk. These are just few of the several results of the COUP published in more than 20 papers. Since COUP gives the most comprehensive view of magnetic activity in young stars ever achieved in the nearest rich cluster of very young stars, it is used here as a touchstone for comparison with NGC\,1893.
% , aimed at the study of the cluster characterization (age, mean absorption, distance) and at the disk properties, by means of infrared excesses.
 
In the present paper, we analyse the properties of the X-ray sources in NGC1893, focusing,  in particular, on the X-ray luminosity of low mass stars. We  compare the X-ray properties of the NGC\,1893 members with infrared excesses (hereafter Class II stars), as defined in \citet{Caramazza_08} and in \citet{Prisinzano_11} with the members without infrared excesses (hereafter Class III stars), that are classified as members of the cluster by means of their X-ray emission \citep[see ][for details]{Caramazza_08, Prisinzano_11}.
The outline of the present paper is the following: in \S \ref{extraction} we describe the data reduction,  the photon extraction procedure, and present our X-ray catalog. In \S \ref{spe_prop} we describe the spectral analysis that we pursued by means of spectral fitting and quantile analysis of X-ray spectra. In \S \ref{spectral_properties} we compare the X-ray properties of Class II and Class III of NGC\,1893 and in \S \ref{variability} we present the variability properties of the members. In \S \ref{ngc1893_x_results} we discuss our results and in \S \ref{summary} we summarize our work.  

\section{Data Reduction}
\label{extraction}
\begin{figure*}[!th]
%  \centering
% \centerline{\psfig{figure=color_true_small_mod.ps,width=8 cm,angle=0}}
% \centerline{\psfig{figure=prova.ps,width=8 cm,angle=0}}
 \centerline{ \includegraphics[clip=true,width=12cm,angle=0]{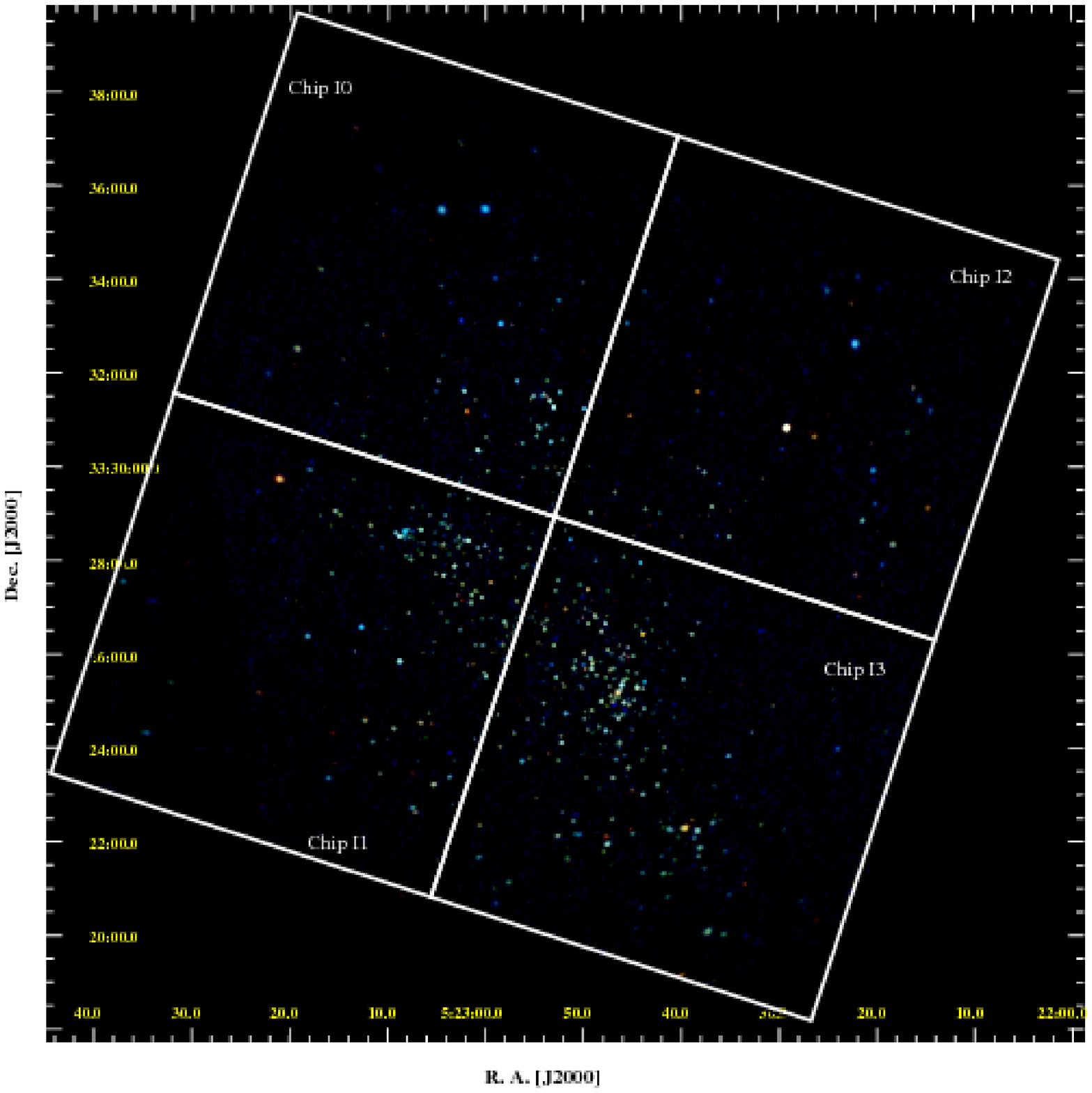}}%generated by /uc3/mcarama/ngc1893_x/analysis/color_image
\caption{True-color image of the young cluster NGC\,1893. A kernel smoothing has been applied to highlight point sources. The energy bands of the RGB image are 0.5-1.1, 1.1-1.9, and 1.9-8.0 kEv for red, green and blue colors. Brightness is scaled to the logarithm of the photon number in the displayed pixel. The color model depicts zero flux as black.}
\label{true_col}
\end{figure*}

The X-ray observations of NGC\,1893 combine four nearly consecutive exposures of the cluster taken in 2006 November and a fifth exposure taken in 2007 January for a total exposure time of $\sim 440$ ks. The combined  X-ray image is shown in Fig. \ref{true_col}. Source detection was performed by \citet{Caramazza_08} and we refer to that work for details of the observations and the detection method. 

Starting from the \citet{Caramazza_08} catalog, we proceeded to the photon extraction of the source photons by means of {\sc ACIS
Extract}\footnote{http://www.astro.psu.edu/xray/docs/TARA/ae\_users\_guide.html}
(AE) v3.131 \citep{acis_extract}, an IDL based package of tools that can assist the observer in performing the many tasks involved in analyzing a large number of point sources observed with the ACIS instrument on Chandra.  AE makes extensive use of TARA \footnote{http://www.astro.psu.edu/xray/docs/TARA/}, a package of  tools in the IDL language for visualizing and analyzing X-ray astronomical data in FITS format , CIAO \citep{Fruscione_06} \footnote{ http://cxc.harvard.edu/ciao/index.html}, the data analysis system written for the needs of users of the Chandra X-ray Observatory , and
FTOOLS\footnote{http://heasarc.gsfc.nasa.gov/docs/software/ftools/}, a general package of software to manipulate FITS files.

The extraction of point sources performed with AE takes into account the Point Spread Function (PSF) for single sources, which strongly depends on the off-axis distance ($\theta$). While the PSF is narrow and approximately circular in the inner part of the field of view ($\theta$\,$\lesssim$\,5'), it has a non-Gaussian shape at large off-axis, becoming broader and more asymmetric.

Moreover AE calculates the shape of the model PSF at each source position (by means of the CIAO task {\sc mkpsf}), considering also that the fraction of the PSF where photons are extracted depends on crowding. In this way, AE finds a compromise between having the largest PSF fraction (thus providing good photon statistics for further spectral and timing analysis) and having the best signal-to-noise ratio, not to mention the importance of avoiding the contamination from nearby sources.

After the calculation of the PSF shape, AE refines the initial source positions, in our case originally estimated by PWDetect \citep{Damiani_97a}
assuming a symmetric PSF, by correlating the source images with the model of local
PSFs.  Following AE science
hints\footnote{http://www.astro.psu.edu/xray/docs/TARA/ae\_users\_guide/node35},
this last procedure was only used for those sources lying  off-axis by greater
than 5 arcmin (321 sources), while for the rest of the sources we
simply adopt mean photon positions. Coordinates listed in Table \ref{tbl:src_properties_main}, as
well as their 1$\sigma$ uncertainties, are the result of this process.% Just the first lines of the table are reported but the entire table is available in electronic form.
% We
% verified that these positions are an improvement over those computed by
% PWDetect, especially at large off-axis angles, comparing the offsets between
% X-ray sources and counterparts in the 2MASS catalog (see \S 4.2).

After re-computing positions, AE defines source extraction regions as polygonal
contours of the model PSF containing a specified fraction of source events
(f$_{\rm PSF}$). Generally, we chose f$_{\rm PSF}$=90\%, and computed the
contours from the PSF for a mono-energetic source with E\,=\,1.49 keV. For a few sources in the denser parts of the field of view this fraction
was reduced so as to avoid contamination with other nearby sources, in the most
extreme cases down to f$_{\rm PSF}$ $\sim$40\%.

Although the ACIS-I instrumental background level is spatially quite uniform,
the actual observed background varies substantially across the NGC\,1893
field due to the extended PSF wings of bright sources and  to their readout
trails. Background was therefore estimated locally for each source, adopting
once again the automated procedure implemented in AE, which defines background
extraction regions as circular annuli with inner radii 1.1 times the maximum
distance between the source and the 99\% PSF contour, and outer radii defined
so that the regions contains more than 100 ``background'' events. In order to
exclude contamination of the regions by nearby sources, background events are
defined from an image that excludes events within the inner
annular radii of all the 1021 sources.

Results of the photon extraction procedure are listed in columns 7-10 of Table \ref{tbl:src_properties_main}, where we give the background-corrected extracted source counts for
0.5-8.0 keV, the associated error and the total background counts expected. In summary, our 1021 X-ray sources span
a wide range of photon flux, from $\sim$3 to $\sim$7100 photons during the exposure time. Most sources are faint (e.g. 60\% have fewer than 50
photons).

In the last six columns of Table \ref{tbl:src_properties_main}, we list some characteristics of the 
sources: the significance (signal-to-noise ratio), Kolmogorov-Smirnov probability that the lightcurves of the sources are constant, the median photon energy, the absorption-corrected X-ray luminosities and a flag indicating the method used to determine them.
\begin{sidewaystable*}
\centering
\caption{NGC1893 {\em Chandra} Catalog:  Basic Source Properties \label{tbl:src_properties_main}}
\Tiny{
\begin{tabular}{rcrrrrrrrrrrrrrc}

\hline
\hline
\\
\multicolumn{2}{c}{Source} &
\multicolumn{4}{c}{Position} &
\multicolumn{4}{c}{Extraction} &
\multicolumn{6}{c}{Characteristics} \\

\multicolumn{2}{c}{\hrulefill} &
\multicolumn{4}{c}{\hrulefill} &
\multicolumn{4}{c}{\hrulefill} &
\multicolumn{6}{c}{\hrulefill} \\

 {Seq. No.} &  {CXOU J} &
 {$\alpha$ (J2000.0)} &  {$\delta$ (J2000.0)} &  {Error} &  {Offaxis angle} &
 {$C_{t,net}$} &  {$\sigma_{t,net}$} &  {$B_{t}$} &  {PSF Frac.} &
 {Signif.} &  {$P_{KS}$} & {Eff. Exp.} &  {$E_{median}$} &  { $\log L_{t,c}$ } &
 {$\log L_{t,c}$ Origin} \\

  {} &   {} &
  {(deg)} &   {(deg)} &   {(arcsec)} &   {(arcmin)} &
  {(counts)} &   {(counts)} &   {(counts)} &   {} &
  {} &   {}   &   {(ks)} &   {(keV)} & {(ergs s$^{-1}$)} & {} \\

  {(1)} &   {(2)} &
  {(3)} &   {(4)} &   {(5)} &   {(6)} &
  {(7)} &   {(8)} &   {(9)} &   {(10)} &
  {(11)} &   {(12)} &   {(13)} &   {(14)} &   {(15)} &   {(16)} \\
\hline
  1 & 052204.16+333353.9 & 80.517357 & 33.564988 & 0.5 & 11.1 & 76.7 & 16.2 & 158.3 & 0.91 & 4.6 & 0.881 & 243.3 & 2.7 & 30.74 & c\\
  2 & 052206.05+333237.7 & 80.525221 & 33.543825 & 0.5 & 10.1 & 48.8 & 13.4 & 111.2 & 0.89 & 3.5 & 0.219 & 245.1 & 3.5 & 30.55 & c\\
  3 & 052207.21+333139.6 & 80.530076 & 33.527692 & 0.5 & 9.5 & 46.8 & 12.7 & 96.2 & 0.89 & 3.5 & 0.403 & 294.7 & 1.6 & 30.41 & c\\
  4 & 052212.21+332845.7 & 80.550897 & 33.47938 & 0.4 & 7.9 & 45.5 & 11.0 & 58.5 & 0.89 & 4.0 & 0.444 & 291.7 & 3.2 & 30.4 & c\\
  5 & 052212.97+333401.1 & 80.554067 & 33.566995 & 0.5 & 9.6 & 42.1 & 13.5 & 112.9 & 0.9 & 3.0 & 0.826 & 332.8 & 3.9 & 30.37 & c\\
  6 & 052214.56+333111.7 & 80.560692 & 33.51992 & 0.3 & 7.9 & 147.0 & 15.0 & 59.0 & 0.89 & 9.5 & 0.0 & 332.8 & 2.2 & 31.08 & b\\
  7 & 052214.84+332907.2 & 80.561862 & 33.485355 & 0.2 & 7.4 & 165.1 & 15.2 & 46.9 & 0.89 & 10.5 & 0.326 & 336.5 & 1.2 & 31.87 & a\\
  8 & 052215.68+333125.7 & 80.565369 & 33.52383 & 0.2 & 7.8 & 191.0 & 16.2 & 50.0 & 0.89 & 11.4 & 0.0010 & 346.8 & 2.8 & 31.23 & a\\
  9 & 052216.36+333140.9 & 80.56818 & 33.52805 & 0.2 & 7.8 & 151.1 & 15.1 & 55.9 & 0.9 & 9.7 & 0.0 & 350.0 & 1.4 & 30.69 & a\\
  10 & 052216.99+333044.3 & 80.570822 & 33.512326 & 0.3 & 7.3 & 85.6 & 12.0 & 42.4 & 0.89 & 6.8 & 0.04 & 347.4 & 1.6 & 31.17 & a\\
  11 & 052218.45+332820.8 & 80.576895 & 33.472458 & 0.1 & 6.6 & 311.8 & 19.0 & 30.2 & 0.9 & 15.9 & 0.128 & 359.1 & 1.4 & 31.13 & a\\
  12 & 052218.98+333250.6 & 80.579105 & 33.547401 & 0.4 & 7.9 & 34.6 & 10.0 & 51.4 & 0.89 & 3.3 & 0.661 & 348.2 & 2.7 & 30.28 & c\\
  13 & 052219.11+332421.4 & 80.579652 & 33.405964 & 0.3 & 7.5 & 62.3 & 11.8 & 59.7 & 0.89 & 5.0 & 0.199 & 348.4 & 2.0 & 30.8 & b\\
  14 & 052219.24+332259.2 & 80.580177 & 33.383129 & 0.4 & 8.3 & 85.2 & 13.1 & 65.8 & 0.9 & 6.2 & 0.038 & 345.9 & 2.9 & 30.97 & b\\
  15 & 052219.30+332904.9 & 80.580423 & 33.484722 & 0.3 & 6.4 & 41.0 & 8.9 & 28.0 & 0.9 & 4.3 & 0.0020 & 357.3 & 2.9 & 30.36 & c\\
  16 & 052219.53+332729.7 & 80.581412 & 33.458257 & 0.3 & 6.4 & 41.0 & 9.1 & 32.0 & 0.9 & 4.2 & 0.3 & 362.0 & 1.5 & 30.36 & c\\
  17 & 052219.52+332754.6 & 80.581338 & 33.465173 & 0.2 & 6.4 & 125.9 & 13.0 & 29.1 & 0.9 & 9.3 & 0.57 & 347.4 & 3.1 & 31.11 & a\\
  18 & 052219.66+333348.5 & 80.581954 & 33.563493 & 0.4 & 8.4 & 62.8 & 12.1 & 66.2 & 0.9 & 5.0 & 0.447 & 343.9 & 2.5 & 31.17 & b\\
  19 & 052219.97+332837.3 & 80.583228 & 33.477032 & 0.4 & 6.3 & 26.4 & 7.7 & 23.6 & 0.9 & 3.2 & 0.544 & 326.4 & 1.6 & 30.16 & c\\
  20 & 052220.26+332912.7 & 80.584438 & 33.486863 & 0.2 & 6.3 & 114.1 & 12.4 & 26.9 & 0.9 & 8.8 & 0.995 & 358.9 & 3.0 & 31.05 & a\\
  21 & 052220.39+332942.7 & 80.584974 & 33.495209 & 0.3 & 6.3 & 56.8 & 9.8 & 28.2 & 0.89 & 5.5 & 0.839 & 362.9 & 3.8 & 30.78 & b\\
  22 & 052220.47+332424.6 & 80.585307 & 33.406844 & 0.4 & 7.2 & 42.9 & 10.0 & 45.1 & 0.89 & 4.1 & 0.017 & 362.2 & 1.7 & 30.38 & c\\
  23 & 052220.46+332955.8 & 80.585257 & 33.498841 & 0.1 & 6.4 & 331.1 & 19.6 & 31.9 & 0.89 & 16.5 & 0.432 & 351.0 & 2.8 & 31.52 & b\\
  24 & 052220.62+333030.6 & 80.585958 & 33.50851 & 0.3 & 6.5 & 56.7 & 9.9 & 29.3 & 0.89 & 5.5 & 0.0090 & 346.4 & 1.6 & 30.16 & b\\
  25 & 052221.54+332851.2 & 80.589788 & 33.480892 & 0.1 & 6.0 & 294.7 & 18.4 & 23.3 & 0.9 & 15.6 & 0.0 & 368.3 & 1.6 & 31.44 & a\\
\hline
\hline
\end{tabular}
}
\note{Table~\ref{tbl:src_properties_main} is available in its entirety in the electronic edition.
\\Col.\ (1): X-ray catalog sequence number, sorted by RA.
\\Col.\ (2): IAU designation.
\\Cols.\ (3) and (4): Right ascension and declination for epoch (J2000.0).
\\Col.\ (5): Estimated standard deviation of the random component of the position error, $\sqrt{\sigma_x^2 + \sigma_y^2}$.  The single-axis position errors, $\sigma_x$ and $\sigma_y$, are estimated from the single-axis standard deviations of the PSF inside the extraction region and the number of counts extracted.
\\Col.\ (6): Off-axis angle.
\\Cols.\ (7) and (8): Net counts extracted in the total energy band (0.5--8~keV) in the extraction region; average of the upper and lower $1\sigma$ errors on col.\ (7).
\\Col.\ (9): Background counts expected in the source extraction region (total band).
\\Col.\ (10): Net counts extracted in the hard energy band (2--8~keV).
\\Col.\ (10): Fraction of the PSF (at 1.497 keV) enclosed within the extraction region. A reduced PSF fraction (significantly below 90\%) may indicate that the source is in a crowded region.
\\Col.\ (11): Photometric significance computed as net counts divided by the upper error on net counts.
\\Col.\ (12): Kolmogorov-Smirnov probability that the lightcurves of the sources are constant.
\\Col.\ (13): Effective exposure time: approximate time the source would have to be observed on-axis (no telescope vignetting) on a nominal region of the detector (no dithering over insensitive regions of the detector) to obtain the reported number of counts.
\\Col.\ (14): Background-corrected median photon energy (total band).
\\Col.\ (15): Absorption-corrected X-ray luminosities.
\\Col.\ (16): Origin of the X-ray luminosity, in col. \ (15): (a) from fit to a one temperature model spectrum; (b) from analysis of
quantiles; (c) from a conversion factor derived from the analysis of the quantiles.}
\end{sidewaystable*}

\section{Spectral Analysis}
\label{spe_prop}
In order to determine the intrinsic luminosity of our sources, spectral analysis is necessary. Unfortunately, spectral fitting can be applied only to high counts sources while relatively faint sources with poor statistics cannot be investigated using this method. 

To overcome this limit, we decided to perform spectral fitting (\S \ref{fit_spectra}) on 311 sources for which it is possible to bin the spectrum so that, in the background-subtracted (net) spectrum, each rebinned channel
achieves a signal-to-noise ratio at least 3. In situations where the background
level is high, this approach can produce higher quality binning compared to fixing a minimum number of counts in the source
spectrum. For less intense sources we analyzed the spectrum of sources considering the quantiles of the distribution of energy for source counts. As we will describe in the following for sources down to 30 counts, applying a quantile method \citep{Hong_04}, we have been able to derive the values of  $N{_H}$ and $kT$ interpolating directly from a thermal grid of models with a good accuracy, while for very low statistics sources we have used the quantile analysis to derive a median conversion factor from the count rate to the X-ray luminosity. Energy quantiles have been also used to discern the extragalactic contaminants.
\subsection{Spectral Fitting}
\label{fit_spectra}
% For sources with more than 60 photons, we were able to fit the spectrum.
Reduced source and background spectra
in the 0.5-8.0 keV band have been produced with AE (see \S \ref{extraction}), along with individual ``redistribution matrices files'' (RMF) and ``ancillary
response files'' (ARF).% For model fitting, spectra have been grouped so that in the background-subtracted spectrum each group achieves at least a specified significance: we selected signal-to-noise equal to 3, this allows us to analyze sources with approximately more than 60 net counts.

We fit our spectra assuming emission by a thermal plasma, in collisional
ionization equilibrium, as modeled by the {\sc APEC} code
\citep{Smith_01a}. Elemental abundances are not easily constrained
with low-statistics spectra and were fixed at Z=0.3 Z$_\odot$ \citep[see][]{Prisinzano_11}, with solar
abundance ratios taken from \citet{Anders_89}. The choice of sub-solar
abundances is suggested by several  X-ray  studies of star forming regions
\citep[e.g.][]{Feigelson_02a, Maggio_07}.  Absorption was
accounted for using the {\sc WABS} model, parametrized by the hydrogen column
density, N$_{\rm H}$ \citep{Morrison_83}.

% \begin{figure}[!ht] \centering
% \includegraphics[width=9cm,angle=0]{fig_8.ps} \caption{Median N$_{\rm H}$
% and kT vs. source counts for sources fit with one-temperature models. Filled
% and empty squares indicate the average values of the $N_H$ and kT in each
% considered count range with vertical scales given on the right and left-hand
% side, respectively. The extent of the count ranges is indicated by the
% horizontal error bars. Vertical error bars indicate the median of the 1$\sigma$
% uncertainties  on the two parameters. For this plot we only considered N$_{\rm
% H}$  and kT values with formal relative errors smaller than 90\%. The hatched
% area below 20 counts indicates the count-range in which spectral fits were
% discarded because of strong biases in the best fit parameters. } \label{xcount}
% \end{figure}

We fit source spectra with
one-temperature (1T) plasma models using an automated procedure. In order to
reduce the risk of finding a relative minimum in the $\chi^2$ spaces,  our
procedure chooses the best fit among several obtained starting from a grid of
initial values of the model parameters: log(N$_{\rm H}$)\,=\,21.0, 22.0 cm$^{-2}$ and kT\,=\,0.5, 1.0, 2.0, 10.0 keV.
% As noted by many authors
% \citep[e.g.][]{2002ApJ...575..354G,2005ApJS..160..319G,2006astro.ph..4243F},
% the spectral fitting of low statistics ACIS sources is problematic  because of a
% degeneracy between plasma temperature and absorption. The degeneracy also
% results in a systematic bias:  kT values are often underestimated by as much
% as  $\sim$50\% while N$_{\rm H}$ values are  overestimated. We investigated
% this issue with our data by considering the distributions of the best fit
% parameters for source in different count-statistics bins. Figure \ref{xcount}
% shows the run of mean kT and N$_{\rm H}$ with source counts for spectra fitted
% with 1T models. The  systematic decrement of N$_{\rm H}$ and the increment of
% kT with increasing source statistics are hardly explainable as physical effects
% and rather indicate that the spectral parameters obtained for sources with less
% than $\sim$20 photons are ill-constrained.

% \input{thermal_spectroscopy}

Among the 311 sources for which we conducted the spectral fitting, for 286 sources it was possible to determine $N_H$ and $kT$.  Using the spectral fit parameters, we have then  derived X-ray luminosities for each of the 286 sources modeled with one-temperature spectrum, considering  for NGC\,1893 a distance of 3.6 kpc \citep{Prisinzano_11}. This sources are tagged with an "a" in column (16) in Table \ref{tbl:src_properties_main} and their luminosity is listed in column (15). The 25 sources for which we were unable to fit the spectra with a one temperature model, have been analysed with the study of the energy quantiles of the spectrum (see following sections).

\subsection{Quantile Analysis}
\begin{figure}[!t]
\centerline{\psfig{figure=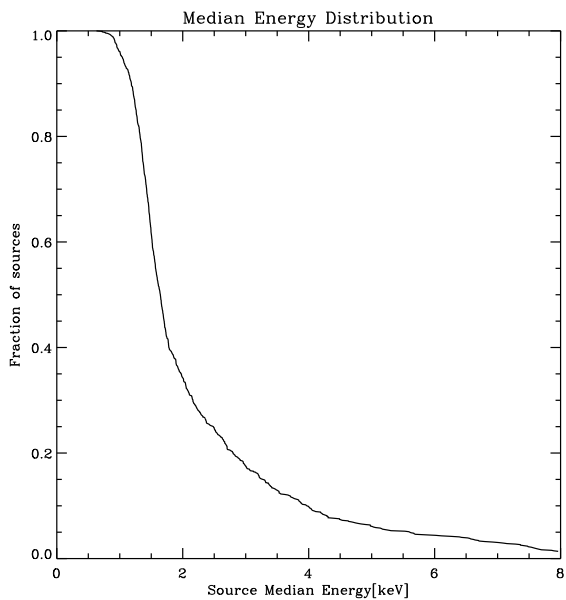,width=8 cm,angle=0}}% file generated by /uc3/mcarama/ngc1893_x/analysis/spectra/plot_spectra.pro
\caption{Cumulative distribution function of the photon median energy of each source computed in the [0.5-8.0] keV range for all the sources. %This kind of non parametric analysis provides global information on the entire sample, allowing to analyze also the sources with poor statistics.
}
\label{med_en}
\end{figure}

Spectral fitting analysis cannot be applied to faint sources, but a rough analysis of the whole sample can be achieved by the study of the median photon energy of each source.  We plot in Fig. \ref{med_en} 
the distribution of the source median energy computed in the [0.5-8.0] keV range. From this plot, we can retrieve some information about the population of our X-ray sources: the median value of the distribution is 1.6 keV, the bulk of the distribution is between 1 and 2 keV (62\% of the median photon energy of the sources lies in this range), while 10\% of the values are over 4 keV; this energetic tail of the distribution is probably due to extragalactic contaminants, even if we cannot exclude at this stage that stellar flares on NGC\,1893 members play some role. While this kind of analysis is very helpful to understand the behavior of the entire sample, it does not help much for the characterization of single sources.

\begin{figure}[!t]
\centerline{\psfig{figure=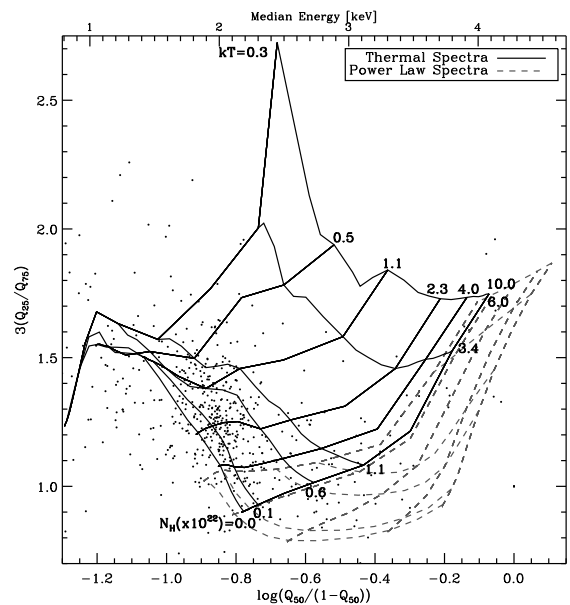,width=8 cm,angle=0}}%file generated by  /uc3/mcarama/NGC\,1893_X/ANALYSIS/SPECTRA/QUANTILE/plot_kt_nh_paper.pro
\caption{Quantiles of X-ray spectra. The dots indicate the quantities derived from source energy quantiles. The two grids shows the same quantities derived from simulated spectra. The solid line grid refer to thermal spectra with $N{_H}=[10^{20}, 10^{21}, 0.6 \cdot 10^{22}, 1.1\cdot 10^{22}, 3.4\cdot 10^{22},6\cdot 10^{22}]$ $cm^{-2}$ and $kT=[ 0.3, 0.5, 1.1, 2.3, 4.0, 10.0]$ keV, while the dashed line grid to simulated power law spectra, typical of AGN, with the same $N{_H}$ values and power law indices $\Gamma=[0.0, 0.4, 1.0, 1.6,2.0]$. }
\label{fig_quant}
\end{figure}
\begin{figure}[!t]
\centerline{\psfig{figure=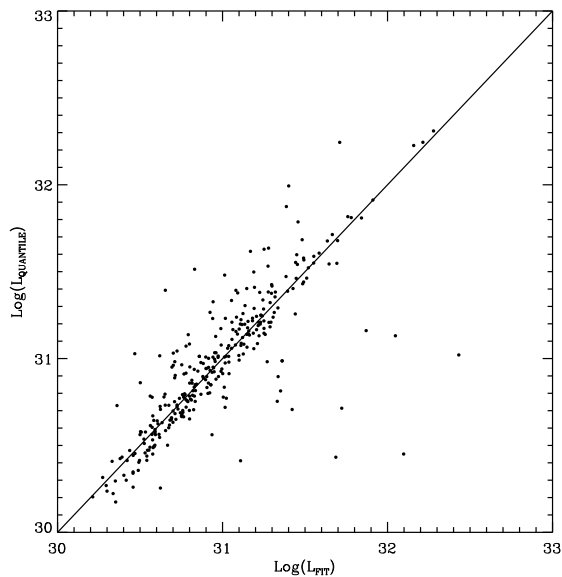,width=8 cm,angle=0}}%file generated by  /uc3/mcarama/NGC\,1893_X/ANALYSIS/SPECTRA/QUANTILE/plot_kt_nh_paper.pro
\caption{Scatter plot of the X-ray luminsity derived from spectral fitting versus that derived from quantile analysis. The solid line indicates equal values of $L_{X}$ derived with the two methods.}
\label{quant_vs_fit}
\end{figure}

 
The conventional spectral classification of weak sources is often performed in the literature by means of  the ratios of source counts in different spectral bands,
i.e. X-ray hardness ratios (XHR)  which can be
considered ``X-ray colors'' \citep{Schulz_89,Prestwich_03}. The flaw in this method is that the choice of the  bands is strongly related to the spectral shape and that, depending on the number of counts in each band, the error bar related to the ratio may be very large. For these reasons, we decide to apply an alternative method, introduced by \citet{Hong_04},  that uses the energy value that divides photons into predetermined fraction, instead of the ratio of counts in prefixed bands. Following  \citet{Hong_04}, if $E_{x\%}$ is the energy below which the net
counts are $x\%$ of the total counts,  quantile $Q_{x}$ is $Q_{x} = \frac{ E_{x\%}- E_{lo}}{E_{up}-E_{lo}}$,
where $E_{lo}$ and $E_{up}$ are the lower and upper boundary of the
full energy band, i.e.  0.5 and 8.0 keV in our case. The fractions we choose are the median (50\% quantile), and the quartiles (25\% and  75\% quantiles). Since quantiles, for a given spectrum, are not independent, \citet{Hong_04} selected the independent variables $log(Q_{50}/(1-Q_{50}))$ and $3(Q_{25}/Q_{75})$, in order to have information on the photon population in several portions of the spectrum. The derived quantile values will be compared with those calculated from simulated spectra in order to derive the best spectral parameters. Figure \ref{fig_quant}
shows the scatter plot of quantiles measured for our sources, compared with grids calculated from simulated spectra. The solid line grid refers to thermal spectra, while the dashed grid to power law spectra,  with spectral index typical of AGNs \citep{Brandt_01}. We note that, as expected, most of the sources lies in the thermal grid region. By interpolating the loci of our sources inside the thermal grid it is possible to attribute values of $N_H$ and $kT$ to each source and derive the intrinsic X-ray luminosity at the distance of NGC\,1893. In figure \ref{quant_vs_fit} we compare, for the brightest sources, the X-ray luminosity derived from the quantile analysis with that derived from spectral fitting. The luminosities derived spectral fitting are generally consistent with those derived from quantile analysis and a linear regression between the two luminosities results in a coefficient $1.04 \pm 0.4$ giving strong support to the effectiveness of the method. The sources for which the luminosity was derived from the quantile analysis are indicated with a "b" in column (16) of table \ref{tbl:src_properties_main}.

By analysing the sources outside the region covered by the thermal grid, we found three subclasses of objects: 
\begin{itemize} 
\item {\it Two temperatures spectra}. There is a small fraction of fairly strong sources (between 30 and 50 photons) that lie over the  left bound of the thermal grid. Due to the concavity of the grid, this is the locus where thermal spectra characterized by two temperatures lie. For these cases, in our one temperature approximation, we associated them to the lowest $N_H$ curve, interpolating just the $kT$ value, that depends basically from the median energy of the spectrum (see Fig. \ref{fig_quant}).
\item {\it  Very poor statistics spectra}. Most of the sources outside the thermal grid are sources with less than 30 photons, with very large errors on quantiles. For these objects, we decided to adopt a single median conversion factor from count rate to luminosity. The conversion factor was derived from the quantile analysis of sources with more than 30 photons and applied to sources with less than 30 photons. These sources are indicated with a "c" in column (16) of Table \ref{tbl:src_properties_main}.
\item {\it Power law spectra}. A small fraction of the sources lie within the power law grid and will be discussed in \S \ref{section_agn}.
\end{itemize}
% First of all, there are the sources lying in the power law grid that will be analysed in \S \ref{section_agn}. 
%  Analysing the sources out of the thermal grid, we found that most of them are very poor statistic sources (less than 30 photons) with very large errors on quantiles. We therefore decided to retrieve a median conversion factor from count rate to luminosity. The conversion factor was derived from the quantile analysis of sources with more than 30 photons and  it to each with less than 30 photons. Nevertheless, we note that there are also relative intense sources (between 30 and 50 photons) that lie over the  left bound of the thermal grid: it is worth noting that, due to the concavity of the grid, thermal spectra characterized by two temperatures can lie in this region. For these cases, in our one temperature approximation, we considered that the median energy has smaller error compared to the ratio of quantiles and shifted them vertically until to the lowest $N_H$ curve, interpolating just the $kT$ value.

\subsubsection{Rejecting extragalactic contaminants}
\label{section_agn}
As shown above, with the energy quantiles analysis it is possible to attribute a luminosity to each source in the sample. It is also useful to investigate the contaminant population in the sample. NGC\,1893 lies toward the galactic anticenter, therefore we expect to detect a large number of extragalactic sources. Moreover, the long exposure necessary to better investigate the faint population of our distant cluster results in a high probability of detecting AGNs. In order to estimate the number of AGNs, we considered the sensitivity for three different off axis regions in the {\it Chandra} ACIS-I field of view and calculated the minimum flux that we expect from a detected  extragalactic source with a typical power law index $\Gamma=1.4$ \citep{Brandt_01} in the direction of  NGC\,1893, \citep[$N_H=6.56 \cdot 10^{21}$ cm$^{-2}$, ][]{Dickey_90}. By comparing the obtained fluxes with the log N-Log S obtained from the Chandra Deep Field North \citep{Bauer_04}, we expect to find $\sim190$ AGNs among our sources. 

From the diagram in Fig. \ref{fig_quant}, we find 128 sources that are outside the thermal grid and seem to be compatible with the power law grid. We have tagged them as candidate AGNs in our catalog, but we have not excluded them from further analysis. Caution is needed because a very hard spectrum, due e.g. to flares can be confused with a power law spectrum; indeed, from the analysis of the variability, we have found that 16 sources tagged as candidate AGN show big flares. %, while 22 of them are classified as Class II stars \citep{Prisinzano_11}}.
%Despite our prudence in excluding possible members of the cluster, 
The spectral analysis finds support from the spatial distribution of candidate AGNs. Looking at Fig. \ref{true_col}, the morphology of the  cluster is evident and it is clear that the distribution of sources in the four instruments chips is not uniform. We compared the number of sources detected in each chip with the number of candidate AGNs.  As we expected, the total number of sources is higher in chips I1 and I3, where the peak density of the cluster lies, while the spatial distribution of the candidate AGNs is different and not related to the spatial distribution of the cluster. Again, we note that the candidate AGNs that show flares follow the same space distribution of the cluster, therefore we treat this population with caution but we do not exclude it from the sample. This analysis is summarized in Table \ref{tab_agn}.
\begin{table*}
\vspace{2.5truecm}
\tabcolsep 0.15truecm
\caption{ Number of detected X-ray sources and AGN candidates in the 4 {\it Chandra} ACIS-I chips.}
\centering
\begin{tabular}{ccc}

\hline
\hline
\\
CHIP & Number of X-ray Sources & Number of AGN Candidates \\
\\
\hline
I0& 158& 28 (0 with flares)\\
I1& 268& 24 (5 with flares)\\
I2& 143& 30 (1 with flares)\\
I3& 452& 46 (10 with flares)\\
\hline
\hline
\end{tabular}
\label{tab_agn}
\end{table*}

\subsection{Upper limit  to the X-ray luminosity of undetected NGC\,1893 members}
The purpose of the present work is to examine the X-ray properties of the NGC\,1893 members, investigate the properties of stars in different evolutionary stages and compare them with those of other star formation regions. For these reasons, we take advantage of the work described in \citet{Prisinzano_11} in which the optical and the infrared properties of NGC\,1893 members are described. In order to compare the luminosity of stars from different classes without any bias due to different depth of infrared, X-ray and optical images, we derived upper limits to the X-ray luminosity for all the cluster members in  \citet{Prisinzano_11} falling in the \textit{Chandra} ACIS-I field of view and undetected in X-rays. Upper limits to the photon count rates were calculated with PWDetect \citep{Damiani_97a,Damiani_97b}, with a detection threshold significance set to 4.6$\sigma$, the same used for source detection \citep{Caramazza_08}. In order to convert count rate upper limits into X-ray luminosity for X-ray undetected members, we calculated a conversion factor, considering the median value of the ratio between X-ray luminosity and count rate for the detected X-ray sources as done for the detected sources. 
In table \ref{tbl:upl_table} we list the sequence number of the source in \citet{Prisinzano_11}, the coordinates and the upper limit to the X-ray luminosity of the 158 Class II members that we will analyse.
 \begin{table*}
\centering
\caption{Upper limit to the X-ray luminosities for NGC1893 X-ray undetected
members \label{tbl:upl_table}}
\Tiny{
\begin{tabular}{rrrr}
\hline
\hline
\\
{ID \citep{Prisinzano_11}} & {$\alpha$ (J2000.0)} &  {$\delta$ (J2000.0)} &  { $\log L_{t,c}$ }
\\
{} & {(deg)} &   {(deg)} & {(ergs s$^{-1}$)} \\

{(1)} &   {(2)} &
  {(3)} &   {(4)}\\
\hline
 408  & 80.562999 & 33.575532 & 30.47\\
 316  & 80.566272 & 33.449992 & 30.31\\
 454  & 80.580917 & 33.562509 & 30.56\\
 123  & 80.580957 & 33.438989 & 30.22\\
 57   & 80.590203 & 33.447045 & 30.29\\
 213  & 80.598221 & 33.455573 & 30.08\\
 492  & 80.598731 & 33.344828 & 30.42\\
 342  & 80.602976 & 33.532386 & 30.24\\
 497  & 80.603157 & 33.522741 & 30.26\\
 458  & 80.606043 & 33.386557 & 30.27\\
 508  & 80.613344 & 33.350895 & 30.35\\
 134  & 80.617675 & 33.454839 & 30.11\\
 319  & 80.618691 & 33.409851 & 30.09\\
 533  & 80.620025 & 33.347069 & 30.61\\
 439  & 80.620694 & 33.359367 & 30.32\\
 164  & 80.624260 & 33.532496 & 30.06\\
 146  & 80.624662 & 33.559724 & 30.16\\
 267  & 80.626424 & 33.450228 & 29.91\\
 445  & 80.628081 & 33.385558 & 30.16\\
 334  & 80.637005 & 33.457892 & 30.09\\
 418  & 80.637332 & 33.361037 & 30.26\\
 93   & 80.640689 & 33.376491 & 30.17\\
 553  & 80.641242 & 33.370493 & 30.2\\
 452  & 80.641437 & 33.378881 & 30.15\\
 422  & 80.642576 & 33.436029 & 29.9\\
 546  & 80.644803 & 33.350252 & 30.67\\
 243  & 80.647512 & 33.363512 & 30.71\\
 459  & 80.648981 & 33.430316 & 29.9\\
 469  & 80.650405 & 33.388107 & 30.06\\
 522  & 80.660049 & 33.399277 & 30.01\\
 575  & 80.662935 & 33.419157 & 29.85\\
 390  & 80.663318 & 33.413863 & 29.88\\
 348  & 80.663623 & 33.396689 & 30.02\\
 394  & 80.671760 & 33.378814 & 30.28\\
 337  & 80.676571 & 33.476855 & 29.7\\
 466  & 80.677781 & 33.395425 & 30.01\\
 506  & 80.679261 & 33.379202 & 30.12\\
 136  & 80.680467 & 33.599258 & 30.31\\
 496  & 80.680484 & 33.396474 & 30.1\\
 544  & 80.681921 & 33.495006 & 29.7\\
 296  & 80.684036 & 33.394367 & 30.18\\
 501  & 80.684164 & 33.391812 & 30.05\\
 381  & 80.684924 & 33.393304 & 30.02\\
 524  & 80.689466 & 33.559017 & 30.39\\
 168  & 80.689949 & 33.412581 & 30.16\\
 476  & 80.690269 & 33.350649 & 30.23\\
 472  & 80.691751 & 33.341261 & 30.28\\
 456  & 80.692317 & 33.399608 & 30.15\\
 480  & 80.693894 & 33.367976 & 30.13\\
 54   & 80.697590 & 33.389405 & 31.02\\
 320  & 80.697977 & 33.395933 & 29.94\\
 434  & 80.698224 & 33.40569 & 31.37\\
 488  & 80.699277 & 33.409286 & 29.88\\
 362  & 80.701203 & 33.354283 & 30.21\\
 165  & 80.702890 & 33.527785 & 30.29\\
 513  & 80.703693 & 33.400595 & 29.95\\
 389  & 80.704304 & 33.416291 & 29.82\\
 427  & 80.706539 & 33.412613 & 29.85\\
 289  & 80.707190 & 33.411752 & 29.86\\
 94   & 80.707273 & 33.428734 & 30.37\\
 413  & 80.707867 & 33.464616 & 29.68\\
 504  & 80.708417 & 33.437063 & 29.71\\
 401  & 80.709488 & 33.513873 & 29.89\\
 437  & 80.710054 & 33.421240 & 30.37\\
 364  & 80.710109 & 33.462006 & 29.71\\
 447  & 80.710368 & 33.467710 & 29.64\\
 195  & 80.711121 & 33.404267 & 29.91\\
 449  & 80.711743 & 33.4208385 & 29.81\\
 64   & 80.712482 & 33.567166 & 30.09\\
 59   & 80.712499 & 33.388352 & 29.98\\

\hline
\hline
\end{tabular}
}
\note{Table~\ref{tbl:upl_table} is available in its entirety in the electronic edition.
\\Col.\ (1): \citet{Prisinzano_11} sequence number, sorted by RA.
\\Cols.\ (2) and (3): Right ascension and declination for epoch (J2000.0).
\\Col.\ (4): Estimated upper limit to the Absorption-corrected X-ray
luminosities.}
\end{table*}

\section{X-ray properties of low mass members}
\label{spectral_properties}
Focusing on the X-ray properties of the NGC\,1893 low-mass members, we started from the member catalog of  \citet{Prisinzano_11}. In particular, we selected all the Class II or Class III candidate members with  mass smaller than 2 $M_\odot$ and whose optical colors are compatible with the cluster locus. In order to have a complete sample of sources, we selected all the stars with mass greater than 0.35 $M_\odot$, the choice of this completeness threshold derives from analysis of the color-magnitude diagram \citep[fig. 11 in][]{Prisinzano_11} and will be confirmed in the following. We also required that the candidate members are inside the \textit{Chandra} ACIS-I field of view. In this way, we identify 591 low-mass members, 307 of which are Class II stars, while 284 are classified as Class III stars. Among the 591 members, 158 of the Class II stars are not detected in the X-ray observations, therefore we have only a determination of the upper limit to their X-ray luminosity. The following analysis is based on the sample of 591 low-mass members.

\begin{figure}[!t]
\centerline{\psfig{figure=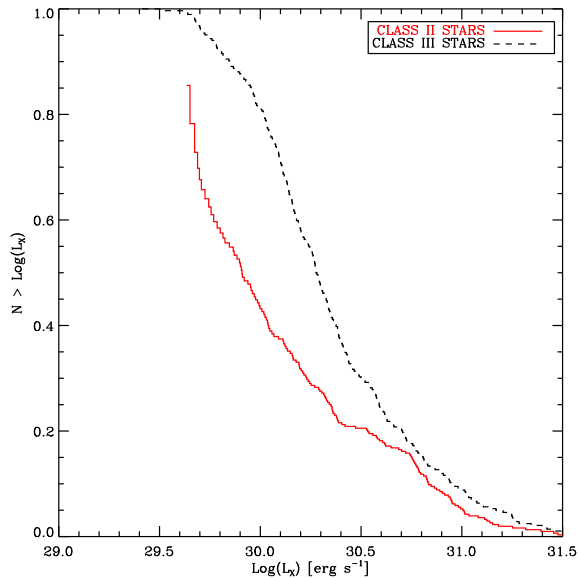,width=8 cm,angle=0}}%file generated by  /uc3/mcarama/NGC\,1893_X/ANALYSIS/SPECTRA/QUANTILE/spectral_properties_paper.pro
\caption{Cumulative X-ray Luminosity Functions of the Class II and Class III members. The nondetections were taken onto account, deriving the XLFs with the Kaplan-Meier maximum likelihood estimator.}
\label{xldf}
\end{figure}

\begin{figure}[!t]
% \centerline{\psfig{figure=lxvsmass.ps,width=8 cm,angle=0}}%file generated by /uc3/mcarama/NGC\,1893_X/ANALYSIS/SPECTRA/QUANTILE/spectral_properties_paper.pro
\centerline{\psfig{figure=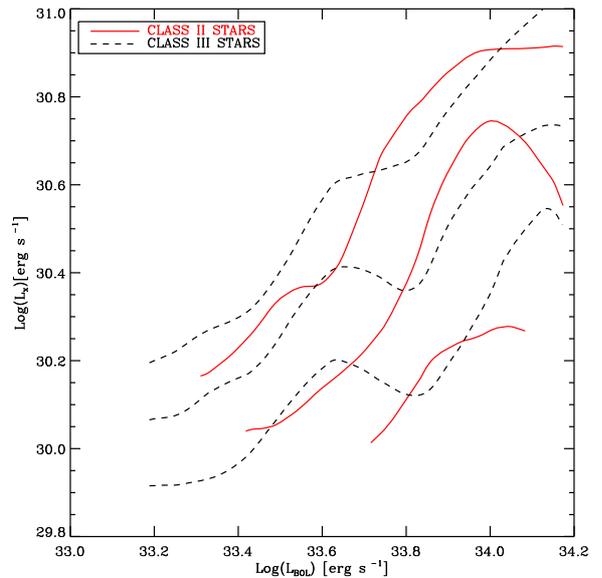,width=8 cm,angle=0}}%file generated by /uc3/mcarama/NGC\,1893_X/ANALYSIS/SPECTRA/QUANTILE/spectral_properties_paper.pro
\caption{X-ray luminosity distribution as a function of the stellar bolometric luminosity. The three couples of lines refer to the (25\%, 50\%, 75\%) quartiles of the ${L_X}$ distributions calculated in running intervals of $L_{BOL}$ constituted by 80 contiguous points and smoothed in interval of bolometric luminosity of 0.1 erg s$^{-1}$. The continuous lines refer to Class II stars, while the dashed ones to Class III stars. The quartiles have been plotted just where the fraction of upper limit values does not affect the calculation of the quartile itself. }
\label{lxvsmass}
\end{figure}

To compare the luminosity of different classes of objects, we analyzed the cumulative  X-ray Luminosity Functions (XLF) of the Class II and Class III members. In order to take into account non-detections,
XLFs were derived using the Kaplan-Meier maximum likelihood estimator \citep{kaplan_meier_58}. Figure \ref{xldf} shows the comparison of the cumulative X-ray luminosity functions of Class II and Class III members. Looking at this global description, we note that the range of luminosity is $3 \cdot 10^{29}-3 \cdot 10^{31}$ $erg$ $s^{-1}$. It is evident that the Class III stars are brighter in X-rays, the median is higher and the body of the Class III object distribution is above that of Class II objects. At this age the dynamo mechanism is saturated in most of the PMS stars, therefore the X-ray luminosity scales as the bolometric luminosity of the star \citep[e.g.][]{Preibisch_05_b}. This means that the difference in the whole XLF may be due to intrinsic properties or to a different bolometric luminosity (or mass) distribution among the two infrared classes. To avoid the mass dependence, it is usual to compare the XLFs in several ranges of mass \citep[e.g.][]{Flaccomio_03,Preibisch_05_b, Prisinzano_08}, despite the large errors caused by having few objects in each bin of mass. We overcome both of these difficulties by analysing the  Class II and Class III X-ray luminosities as a function of the bolometric luminosity of the star. Figure \ref{lxvsmass} shows the quantiles (25\%, 50\%, 75\%) of the $L_X$ distribution in running intervals of bolometric luminosity covering 80 datapoints, for both of Class II (continuous line) and Class III (dashed line) stars. The resulting lines have been smoothed over scales of 0.1 in logarithm of  bolometric luminosity. Note that for Class II stars the quartiles have been plotted only where the fraction of $L_X$ upper limits does not affect the calculation of the quartiles themselves (e.g., the first quartile may be calculated only if  $L_X$ upper limits are less than 25\%). The three quartiles give us different information about the distribution of $L_X$: the first quartile of Class III objects is similar to that correspondent to Class II stars, but we are able to calculate this value only for a small fraction of Class II member, due to the high fraction of  $L_X$ upper limits in the faint sample. The median and the third quartile, that describe the body and tail of the X-ray distributions respectively, show the same feature. They are lower for Class II members but the difference decreases with bolometric luminosity and for bright stars the quartiles are similar to those of Class III members.  


\begin{figure}[!t]
% \centerline{\psfig{figure=lxlbolvsmass.ps,width=8 cm,angle=0}}%file generated by  /uc3/mcarama/NGC\,1893_X/ANALYSIS/SPECTRA/QUANTILE/spectral_properties_paper.pro
\centerline{\psfig{figure=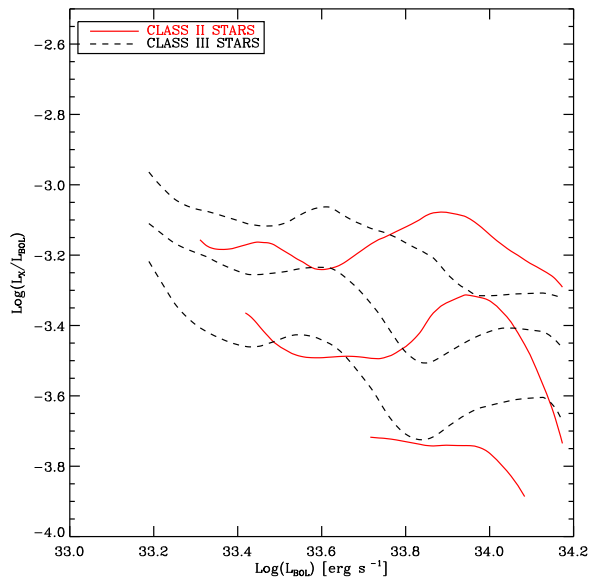,width=8 cm,angle=0}}%file generated by  /uc3/mcarama/NGC\,1893_X/ANALYSIS/SPECTRA/QUANTILE/spectral_properties_paper.pro
\caption{$L_X/L_{BOL}$ distribution as a function of $L_{BOL}$. The three couples of lines refer to the (25\%, 50\%, 75\%) quartiles of the $L_X/L_{BOL}$ distributions calculated in running intervals of  constituted by 80 contiguous points  and smoothed in interval of bolometric luminosity of 0.1 erg s$^{-1}$. The continuous line refer to Class II stars, while the dashed one to the Class III stars.}
\label{lxlbolvsmass}
\end{figure}

In order to interpret these results, we also need to compare the value of $L_X/L_{BOL}$ for the two infrared classes of objects. We show in Fig. \ref{lxlbolvsmass} the three running quartiles of the distribution of $L_X/L_{BOL}$ for Class II and Class III objects. We see that the ratio decreases for the two classes as a function of  $L_{BOL}$. That means that the fraction of stars in the saturated dynamo regime ($L_X/L_{BOL}  \sim 3$) decreases as a function of $L_{BOL}$, and the displacement of Class II $L_X/L_{BOL}$ below that for Class III stars means that the fraction of unsaturated X-ray sources is larger for disked stars compared to diskless stars. 

We also compared the global X-ray properties of the stars in NGC\,1893 with those of the Orion Nebula Cluster, in particular those obtained from the Chandra Orion Ultradeep Project (COUP) \citep{Getman_05}. As we discussed in \S \ref{intro_x}, COUP gives the most complete set of studies in X-rays for young stars ever achieved for such a rich cluster therefore it has been considered a touchstone and its results has been compared with previous and succeeding X-ray studies. \citet{Feigelson_05b} compared the COUP XLF with that of NGC1333 and IC 348 finding that "the shapes of different YSC XLFs appear to be remarkable similar to each other, once a richness-linked tail of high luminosity O stars is omitted ". Also RCW38, IC1396N, NGC 6357, the Carina Nebula Cluster, M17 and NGC2244 \citep{Wolk_06, Getman_07, Wang_07, Sanchawala_07, Broos_07,Wang_08}  compared the X-ray properties of several clusters to COUP results, demonstrating that in the Sun neighbourhood the X-ray properties of star forming regions are similar, comparing clusters of similar age and accounting for the IMF. The only relevant difference in the XLF is for Cep OB3b \citep{Getman_06}, where the XLF has a different shape from that seen in the ONC with an excess at ${\rm Log(L_X)\sim 29.7}$ or 0.3 M$_{\rm \odot}$. The origin of the difference it is not clear but could be ascribable to a deviation in the IMF or some other cause, such as sequential star formation generating a non-coeval population. 

\begin{figure}[!t]
\centerline{\psfig{figure=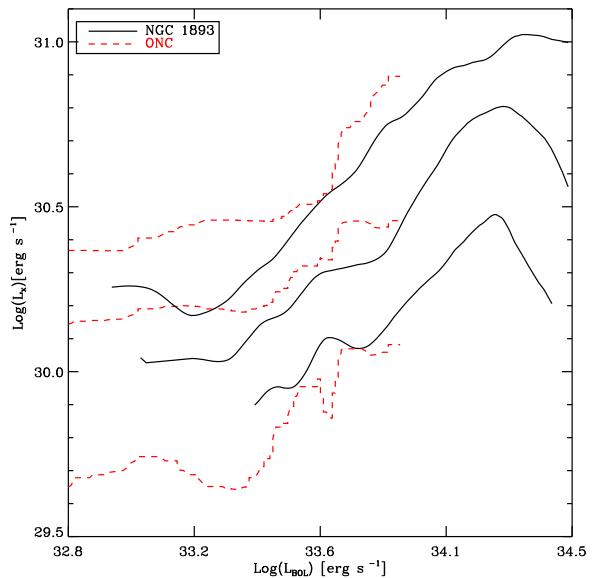,width=8 cm,angle=0}}%file generated by  /uc3/mcarama/NGC\,1893_X/ANALYSIS/SPECTRA/QUANTILE/spectral_properties_paper.pro
\caption{X-ray luminosity distribution as a function of the bolometric luminosity. The three pairs of lines refer to the (25\%, 50\%, 75\%) quartiles of the $L_X$ distributions calculated in running intervals of bolometric luminosity constituted by 80 contiguous points  and smoothed in intervals of bolometric luminosity of 0.1 erg s$^{-1}$. The continuous line refer to the NGC\,1893 stars, while the dashed one to the ONC stars.}
\label{lxvsmass_orion}
\end{figure}

Starting from these previous results, we consider the ONC as representative of the X-ray properties of nearby star forming region and will compare COUP results with those of  NGC\,1893, that is $\sim$10 times more distant from the Sun. Figure \ref{lxvsmass_orion} describes the quantiles of the $L_X$ distributions for the Class II and Class III stars between 0.35 and 2 $M_{\odot}$ in the case of NGC\,1893 (continuous line) and ONC (dotted line). The first quantile of the NGC\,1893 $L_X$ distributions is similar to the ONC, and, since we know that the COUP sample is complete in this range of masses \citep{Getman_05}, it derives  that our sample of stars has the same completeness of that of the ONC. The median of the two $L_X$ distributions are similar, but we note that the third quartile is larger for the ONC at the faintest $L_{BOL}$. This means that we lack bright X-ray sources at the lowest masses in NGC\,1893 and this cannot be related to an incompleteness of the sample: in that case we would loose faint stars and the effect would be enhanced. \citet{Albacete_07b}, comparing the X-ray properties of young stars in Cyg OB2 with those in the ONC, noted that dividing the COUP observation in 100 ks segments, a significant fraction of the longer and more energetic flares are missed. The difference between the two distributions could be therefore explained referring to the fact that COUP observation is longer than NGC\,1893 ones, therefore in the ONC it is possible to observe rare very energetic flares that, even not modifying the median of the distribution, affect its tail. In order to take into account the effect of the duration of the observation, in the comparison of flares properties  between NGC\,1893 and the ONC (see \S \ref{variability}), we will normalize for the exposure time.

\section{X-ray Variability}
\label{variability}
\begin{figure}[!t]
\centerline{\psfig{figure=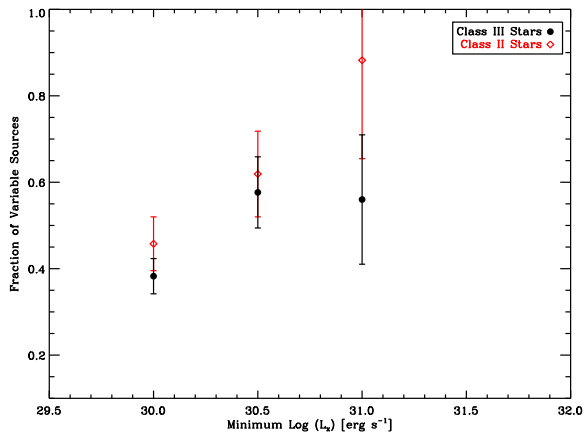,width=8 cm,angle=0}}
\caption{Number of variable stars as a function of the minimum X-ray luminosity of the stars in the subsample. The vertical bars are the standard deviations ${\rm(\sqrt{\rm{ Number\ of\ variable\ stars}}/ Number\ of\ stars\ in\ the\ subsample)}$. }
\label{ks_stat}
\end{figure}

%We studied the variability of our sample, taking account also of different subsamples derived from the infrared classification \citep{Caramazza_08}.
% We  present here the analysis of  the light curves of our X-ray sources and, in particular, those of Class II and Class III NGC\,1893 members. 
We have analyzed the variability of the lightcurves of the X-ray sources, applying the Kolmogorov-Smirnov test
to the arrival times of photons, and determined that 34\% of our 1021 X-ray sources (including both members and non-members) are variable with a probability greater than 95\%. If we restrict ourselves to only members, the variability fraction is 36 \%. When split according to whether a disk is present of not, 34\% of the Class III stars are variable vs. 41 \% of the Class II's.  While we cannot state that these two numbers are statistically different, we have to take into account that Class II are intrinsically fainter than Class III stars and therefore the comparison has to be done analysing sources with the same statistics. Figure \ref{ks_stat} shows the number of variable stars as a function of the minimum X-ray luminosity of the stars in the subsample. We note that, considering just the brightest stars, the discrepancy between Class II and Class III seems to be larger, suggesting a real difference in the fraction of variable stars in the two classes. 
It is interesting to investigate the behavior of Class II stars that still show accretion phenomena and those that do not present evidence of strong accretion. In the \citet{Prisinzano_11} catalog there is a subsample of Class II stars for which $H_\alpha$ photometry is available. Using that tracer \citep[see \S 5.3 in][for details on selection]{Prisinzano_11}, we singled out 50 Class II X-ray emitters with ongoing accretion, as indicated by $H_\alpha$ emission, and 77 that do not show strong $H_\alpha$ emission. The two samples of Class II X-ray emitters show the same fraction of variable light curves.

A summary of the fraction of variable stars for the different subsamples of objects is given in Table \ref{tab_var}.

The Kolmogorov-Smirnov test gives just a global indication of the variability of sources, but does not specify the nature of the variability. We therefore  further  analyzed the lightcurves in order to study the variability due to stellar flares.  To single out flares in the lightcurves, we mapped the lightcurves using Maximum Likelihood
Blocks (MLBs) \citep{Wolk_05, Caramazza_07}. The main characteristic of MLBs is
that, being computed from the photon arrival times, their temporal length is not
based on an {\it a priori} choice of temporal bin
length, but
depends on
the light curve itself; for this reason MLBs are a useful tool to quantify
 different levels of emission, and in particular to detect short impulsive events,
 that
 might be missed by binning the lightcurves using fixed length bins. Applying the same operational definition of flares described in \citet{Caramazza_07} that takes into account both the amplitude and time derivative of the count-rate, we singled out flares in the lightcurve as a sequence of blocks with a high count rate and a high rate of
variation of the
photon flux. We estimated the luminosity of each flare, scaling the total luminosity of the source during the observation by means of a count rate  to luminosity conversion factor; we then computed the energy of each flare, E$_{\rm flr}$, by multiplying the flare luminosity for the flare duration, calculated as the total temporal length of the blocks
associated with the flare. The results of this analysis are summarized in Table \ref{tab_var}, where we state the number of flares per source with energy larger than $2 \cdot 10^{35} $ erg (this choice will be justified in the following). 

\begin{table*}
\vspace{2.5truecm}
\tabcolsep 0.15truecm
\caption{X-ray variability for different subsamples of X-ray sources}
\centering
\begin{tabular}{cccc}

\hline
\hline
\\
Sample & Number of sources & Fraction of variable sources (KS Test 95\%) &Number of flares per source \\
\\
\hline
All X-ray sources& 1021& 0.34 $\pm 0.02 $ & 0.16 $\pm 0.02 $\\
Class III X-ray sources& 285& 0.34 $\pm 0.03 $& 0.19 $\pm 0.02 $\\
Class II X-ray sources& 149 & 0.41 $\pm 0.05 $& 0.27 $\pm 0.04 $\\
Accreting Class II X-ray sources& 50 & 0.40 $\pm 0.09 $& 0.30 $\pm 0.08 $ \\
Not accreting Class II X-ray sources& 77 & 0.41  $\pm 0.07 $& 0.27 $\pm 0.06 $\\
\hline
\hline
\end{tabular}
\label{tab_var}
\end{table*}

\begin{figure}[!t]
\centerline{\psfig{figure=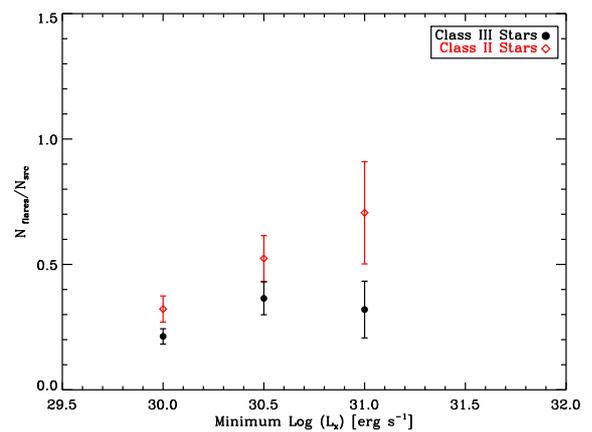,width=8 cm,angle=0}}
\caption{Number of flare per source as a function of the minimum X-ray luminosity of the stars in the subsample. The vertical bars are the standard deviations as in Fig. \ref{ks_stat}.}
\label{flares_stat}
\end{figure}

In Fig.\ref{flares_stat}, we show the number of flares per source as a function of the minimum X-ray luminosity of the analysed subsample. \citet{Caramazza_07} demonstrated that the sensitivity of flare detection methods depends on source statistics, therefore it is interesting to compare the mean number of flares per source in several subsamples of sources, comprising sources with increasing luminosity. We note that for bright sources where we are able to detect all the flares, Class II stars are flaring significantly more than Class III stars and therefore that the disk plays an important role in generating this kind of high energy events.

\begin{figure}[!t]
\centerline{\psfig{figure=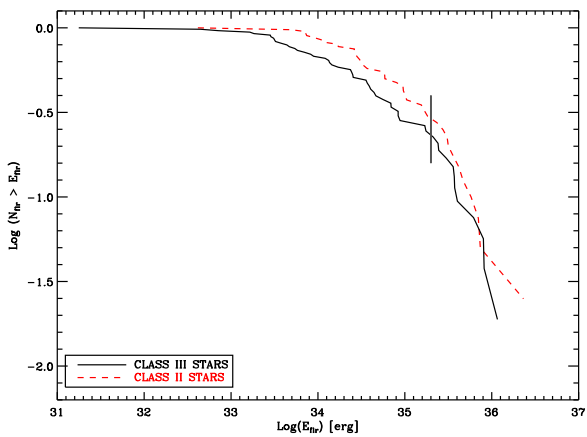,width=8 cm,angle=0}}%file generated by  /uc3/mcarama/NGC\,1893_X/ANALYSIS/LIGHT_CURVE/intensity_distribution_en_paper.pro
\caption{Normalized cumulative distribution function of flare energies for Class III  (solid line) and  Class II stars (dashed line). In the range of energies above the
cutoff value $E_{cut}=2 \cdot 10^{35}$ erg (vertical segment), the two distributions are compatible with a power law with $\alpha-1=1.4\pm 0.4$; for low counts
      the distributions flatten, most likely because the detection of low-counts flares is
      incomplete.}
\label{flare_distr}
\end{figure}

 According to the microflares hypothesis, originally proposed for the solar case \citep[see][]{Hudson_91}, the
X-ray emission of a star can be described as an ensemble of flares with a power law energy distribution:
 \begin{equation} \label {eq: distrib2}
\frac{dN}{dE_{flr}}= k\cdot E_{flr}^{-\alpha}\qquad \textrm{with} \qquad
\alpha>0
\end{equation}
where $N$ is the number of flares with energies between $E_{flr}$ and $E_{flr}+dE_{flr}$, emitted in a
given time interval.
If the index of the power law ($\alpha$) is larger than $2$ even very high levels of apparently quiescent coronal X-ray emission
can be obtained from the integrated effects of many small flares.
Figure \ref{flare_distr} shows the cumulative
 distribution function (continuous line) of the intensity of flares for Class II and Class III members.
For high counts the distribution is well described by a power law, but it
progressively
flattens towards low energies, likely because not all low-count flares are individually detectable.\\
Following the microflares hypothesis,  %(see eq. \ref{eq: distrib2})
we
described the
high energy part of the differential
distribution
of flare counts as a power law with index $\alpha$. 
The cumulative distribution is then described by a power law with index $\alpha-1$.

We have determined the cutoff energy $E_{cut}$ above which the observed distribution
is compatible with a power law and the relative index $\alpha-1$, with the same
method used by \citet{Stelzer_06} \citep[see also ][]{Crawford_70} :
$$\alpha -1=1.2 \pm 0.2     \qquad \qquad E_{cut}=2 \cdot 10^{35} \\ erg$$
In agreement with our assumption on power law shape for the distribution, taking $E_{cut}$
larger than the chosen cutoff, the best fit value of $\alpha$ remains stable within the
uncertainties, while under this threshold we cannot neglect the incompleteness
effect.

Figure \ref{flare_distr} also shows the cumulative distribution of flare energies for the Class II and Class III samples. We noticed that Class II star are more variable and more flaring than Class III objects, but we now see that the flares of Class II and Class III stars show the same distribution. We calculated the slope of the two power laws, finding for the two subsamples $\alpha -1=1.4 \pm 0.4$. 

% \begin{figure}[!t]
% \centerline{\psfig{figure=beta_ks.ps,width=8 cm,angle=0}}%file generated by  /uc3/mcarama/NGC\,1893_X/ANALYSIS/LIGHT_CURVE/intensity_distribution_en_paper.pro
% \caption{}
% \label{beta_ks}
% \end{figure} 

\begin{figure}[!t]
\centerline{\psfig{figure=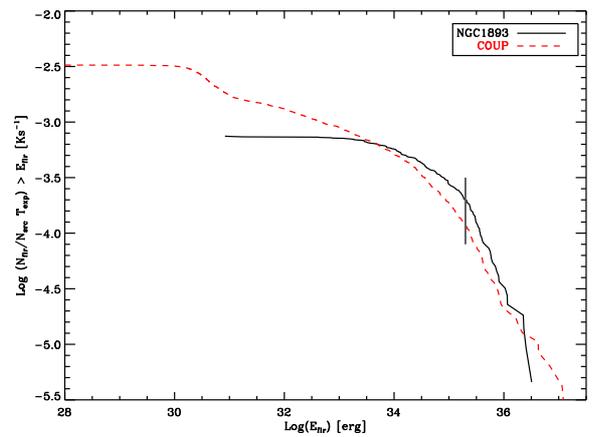,width=8 cm,angle=0}}%file generated by  /uc3/mcarama/NGC\,1893_X/ANALYSIS/LIGHT_CURVE/intensity_distribution_en_paper.pro
\caption{Cumulative distribution function of flare energies for X-ray members in NGC\,1893 (solid line) and ONC (dotted line). The distributions are normalized to the total number of sources of each sample and to the total exposure time. In the range of energies above the
cutoff value $E_{cut}=2 \cdot 10^{35}$ erg (vertical segment), where both the flare energy distributions are complete, the distributions of the flares follow similar distributions.}
\label{flare_distr_coup}
\end{figure}

Figure \ref{flare_distr_coup} shows the comparison between the cumulative distribution function of flare energies for NGC\,1893 (solid line) and the ONC (dotted line). The distributions are normalized to the total number of sources of each sample and to the total exposure time. Above the completeness cutoff energy the two distributions are similar.

\begin{figure}[!t]
\centerline{\psfig{figure=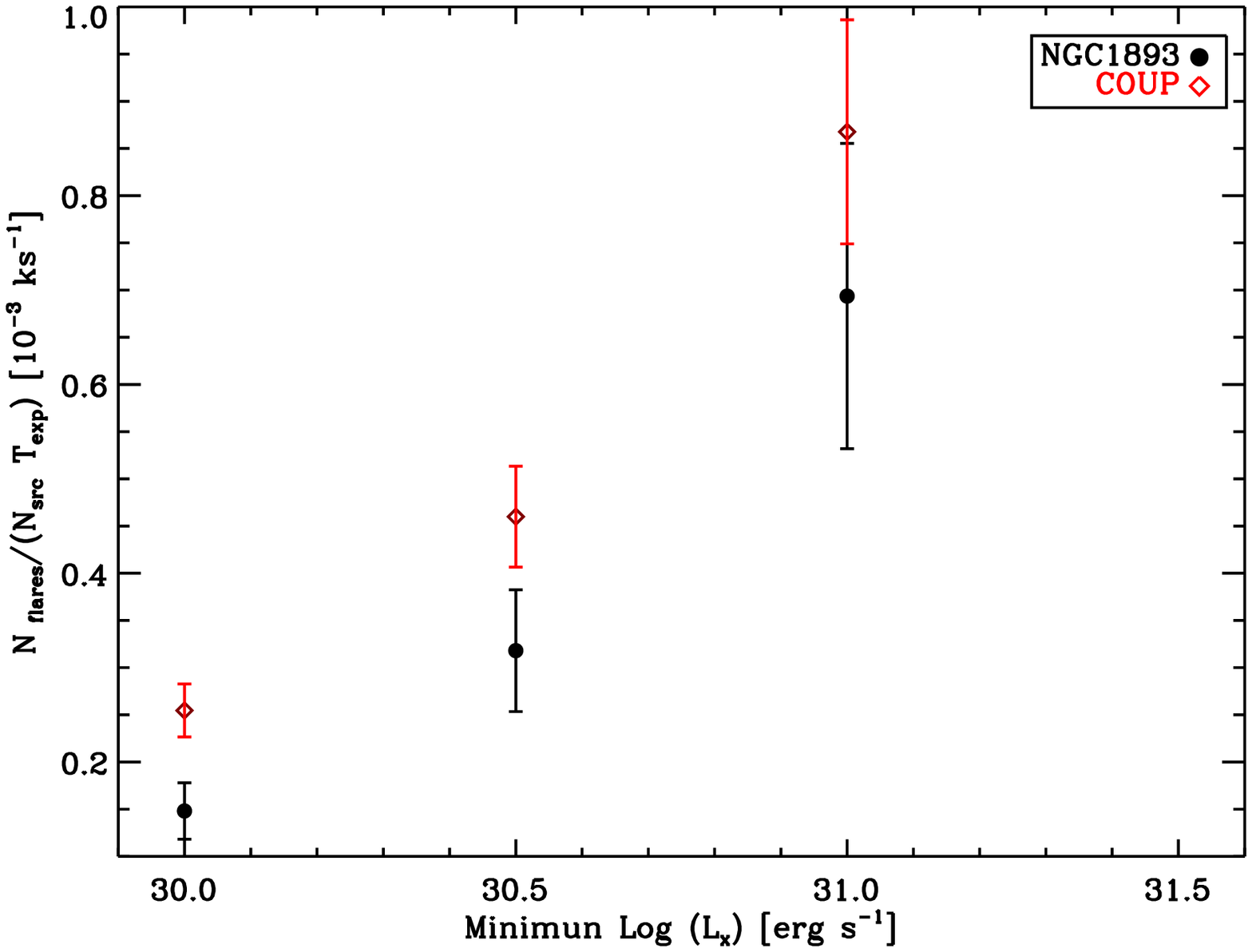,width=8 cm,angle=0}}%file generated by  /uc3/mcarama/NGC\,1893_X/ANALYSIS/LIGHT_CURVE/statistic_histogram_paper.pro
\caption{Mean number of flares per kilosecond above the cutoff value $E_{cut}=2 \cdot 10^{35}$ erg for subsamples of X-ray sources with $L_X$ larger than a given value. Circles refer to the NGC\,1893 X-ray members, while diamonds to the ONC members. The vertical bars are the standard deviations as in Fig. \ref{ks_stat}.}
\label{flare_frequences}
\end{figure}

Figure \ref{flare_frequences} shows the comparison between the mean number of flares for the NGC\,1893 and the ONC samples. In order to compare the same kind of flares, we considered only the flares over the completeness cutoff energy. We note that the number of flares per source is higher in the ONC but the difference decreases when we exclude low statistic sources. We infer that the difference may be due from a likely incompleteness of  the NGC\,1893 sample at low luminosities.

\section{Discussion}
\label{ngc1893_x_results}
The influence of the environment on star formation process may be investigated also through the analysis of X-ray emission.
% For instance, chemical composition is a fundamental ingredient for models of stellar structure since it significantly impacts the opacity of the plasma. Moreover, differences in the mass distribution in a cluster or different disk fractions may affect the global X-ray properties of the cluster itself.  
NGC\,1893, for its position in the Milky Way is a good target to investigate possible differences in the  behavior of stellar coronae and compare stars in clusters at the periphery of the Galaxy with those lying in dense spiral arms in the solar neighborhood. 

We analyzed the X-ray properties of NGC\,1893, in particular comparing the XLF of low mass ($<2 M_{\odot}$) Class II and Class III young stellar objects. We confirmed the known result that the Classical T Tauri Stars (CTTS), i.e. the stars with signatures of disks, are globally less active than the Weak-lined T Tauri stars (WTTS), i.e. stars that have no signatures of optically thick disks \citep[e.g.][]{Preibisch_05_b}, while being more variable. We further investigated this result, analysing the behavior of $L_X$ as a function of the bolometric luminosity.  The correlation between these variables is compatible with the well known saturated dynamo scenario, typical of pre-main sequence stars. Using the method described in \S \ref{spectral_properties}, we demonstrated that, even considering stars with similar $L_{BOL}$, the median  $L_X$ is lower for Class II stars, even if the difference decreases for stars with high $L_{BOL}$. Our statistical analysis suggests that the fraction of unsaturated Class II members is larger than for the Class III members. This result is in agreement with the scenario in which accretors stars are less luminous than non-accreting stars \citep{Stelzer_01, Flaccomio_03b, Stassun_04_a, Preibisch_05_b,Flaccomio_06, Telleschi_07}.  However the reason why Class II should be underluminous in X-rays is not well understood. There  are four main ideas that have been formulated to explain the  different behaviors of disked and diskless stars; two are related to the presence of disk itself and two to the presence of accretion from the disk: 1) The lower X-ray luminosity of Class II stars could be related to the higher extinction due to X-ray absorption by circumstellar disks, even if the COUP survey results do not support this idea \citep{Preibisch_05_b}. 2) A possible explanation is related to the magnetic braking: indeed there are examples of magnetic connection between the star and the disk leading to the idea that disked stars are on average slower than diskless stars \citep{Favata_05, Rebull_06, Prisinzano_08} and have a weaker dynamo action with consequent lower X-ray emission.  In this case, the dynamo process would be the same for the star with and without disk and the lower luminosity of CTTS could be attributed to the slower rotation of CTTS related to the presence of the disk. 3) The presence of accretion may alter the stellar structure and change the dynamo process itself \citep{Preibisch_05_b}, in this case the dynamo process could be different with different  saturation limit for CTTS and  WTTS. 4) Finally, \citet{Gregory_07} demonstrated that the corona is not able to heat all the material in the accretion columns and this cooler  material, not visible in X-rays  may obscure the line of sight to the star, reducing the X-ray emission. Recently, this scenario was also supported by \citet{Flaccomio_10} who found a correlation between the X-ray and optical variability for CTTS, while they did not find any correlation for WTTS. With our sample of data, we are not able to prefer one of the previous hypothesis, but we can add the observational constraint that the difference between the two classes of object is more evident for low mass stars and we can infer that this can be related to the different evolution time of disks around low mass stars and solar mass stars, causing there to be a larger fraction of disked stars at the lowest masses \citep{Carpenter_06, Sicilia_06, Lada_06,Megeath_05}.

We have also compared the X-ray properties of NGC\,1893 with those of the ONC low mass members.  At this age the dynamo mechanism is saturated for a large fraction of the  stars and this leads to a dependence of $L_X$ on the bolometric luminosity \citep[e.g.][]{Preibisch_05_b}. Since we cannot {\it a priori} assume that the mass distribution in the two clusters is similar, we compared the two samples as a function of the bolometric luminosity. With this method, we obtained that the median of the two $L_X$ distributions vs mass are similar, but the third quartile is larger for the ONC at the faintest $L_{BOL}$. This effect seems to indicate an intrinsic lack of bright X-ray sources at the lowest masses in NGC\,1893. This difference with the ONC cannot be an indication of an incompleteness of the  NGC\,1893 sample because, in that case, we would lose faint stars and the effect would be in the opposite direction. This difference between the two clusters in the high luminosity tail of the $L_X$ distributions could be ascribed to the longer duration of COUP observation: indeed, in the ONC there was the possibility to observe long energetic flares that enhance the third quartile of the distribution.%, that is the corner stone for nearby star forming regions.

Analyzing the variability of Class II and Class III stars in our sample, we find the known results that Class II are more variable than Class III stars and show more flares. The explanation of this behavior can be related to the effect of accretion. \citet{Flaccomio_10} observed that the X-ray variability of disked stars can be due to the shielding of significant fraction of the coronal plasma by dense accretion streams of cold material \citep[see][]{Gregory_07}. It is difficult to compare in detail the variability of NGC\,1893 sources with the Orion Nebula Cluster ones, because of the several biases due to the different distance and the different duration of the X-ray observations that lead to a different completeness of the samples.  Taking into account these several biases we find that the X-ray flares in Orion and NGC\,1893 show a similar energy distribution frequency, leading to indistinguishable behavior of the two clusters from the coronal variability point of view. 

We conclude that despite its peculiar location in the Galaxy, NGC\,1893 includes a rich population of Class II and Class III X-ray sources that have, both from the X-ray luminosity and X-ray variability point of view, similar properties to nearby star forming regions, such as the ONC. The fact that the X-ray properties of clusters in such different environments are so similar gives strong support to the \citet{Feigelson_05b} suggestion that the X-ray properties can be used as standard candles, providing a new instrument for measurement for distances of young clusters \citep{Kuhn_10}.

\section{Summary and Conclusion}
\label{summary}
As part of the large multiwavelenght project \textit{The Initial Mass Function in the Outer Galaxy: the star forming region NGC\,1893}, we have analyzed the 450 ks Chandra Observations, and studied the X-ray properties of the 1021 detected sources.  The X-ray data have been combined to optical, near and mid-infrared data \citep{Prisinzano_11} in order to correlate the X-ray properties to the presence of disk and/or accretion. Below, we summarize the overall results of our investigations.
\begin{itemize}
\item{We derived the X-ray luminosity for our 1021 sources taking advantage of spectral fitting for 311 X-ray bright stars and from quantile analysis of the other sources, finding a range of X-ray luminosity $10^{29.5}-10^{31.5}$ erg s$^{−1}$. We have  analyzed, in particular, the low mass stars X-ray luminosity taking into account the classification of candidate members in Class II and Class III members based on infrared excesses of  \citet{Prisinzano_11}. Class III stars  appear intrinsically more
X-ray luminous than Class II stars, even comparing stars with the same bolometric luminosity. This effect may be due to the presence of the "magnetically connected" disk itself or to the ongoing accretion from the disk to the star. }
\item{We have evaluated the variability of X-ray lightcurve using the Kolmogorov-Smirnov test, finding that 34\% of the sources appear to be variable. We have also searched for flares in our lightcurves finding $0.16$ flares per source. We found that Class II stars are more variable and more flaring than Class III stars, while the flare energy properties are the same. This property may be related to the accreting cold material that obscures part of the X-ray emitting material, making accreting stars X-ray lightcurves more variable than the ones of diskless stars but also suggests that the presence of a disk plays a role in generating high energy events. }
\item{Comparing the X-ray properties of NGC\,1893 with those of the nearby star forming region ONC, we find that the X-ray properties in NGC\,1893 are not affected by the environment and that a  stellar population in the outer Galaxy may have the same coronal properties of nearby star forming regions. This provides a strong evidence of the universality of the X-ray properties, and, as a consequence, a useful tool to determine properties, such as the distance or the total population, of young clusters.}

\end{itemize}

\begin{acknowledgements}
 This research has made use of data obtained from the Chandra X-ray Observatory and software provided by the Chandra X-ray Center (CXC) in the application packages CIAO, ChIPS, and Sherpa.
We acknowledge financial contributions from ASI-INAF agreement I/009/10/0, from European Commission (contract N. MRTN-CT-2006-035890) and PRIN-INAF (P.I. Lanza). S.J.W. is supported by NASA contract NAS8-03060 (Chandra). 
We thank the referee for helpful suggestions and comments that improved this work.\\
\end{acknowledgements}

\bibliographystyle{aa}
\bibliography{biblio}

\end{document}